# Modelling the, 0.6 – 0.7, power law of permittivity and admittance frequency responses in random R-C networks.

Baruch Vainas[a]

[a] Soreq Nuclear Research Center, Yavne 81800. Israel

Abstract

The dielectric response of complex materials is characterized, in many cases, by a similar power law frequency dependence of both the real and the imaginary parts of their complex dielectric constant. In the admittance representation, this power law is often shown as the constant phase angle (CPA) response. Apparently, the power that characterizes many different systems, when expressed as the frequency dispersion of conductivity (the real part of admittance) is often found to be in the range of, 0.6-0.7, or having frequency independent, constant phase angles (CPA) of about 54 - 63 deg. The model suggested here is based on series-parallel mixing of resistors' and capacitors' responses in a random R-C network. A geometric mean evaluation of the effective resistivity of conductors having a uniform distribution of resistivity is used. In contrast to models based on percolation arguments, the model suggested here can be applied to both 2D and 3D systems.

## Introduction

It has been shown that the frequency response of random R-C networks having a binary composition of elements can model the response of systems composed of conductive and capacitive spatial domains. For example, 1 Kohm resistors and 1 nF capacitors positioned at random locations on square R-C networks (Fig. 1) have a well defined frequency region (Fig. 2) where the phase angle between the applied AC current and the corresponding voltage drop has a constant value which is independent of frequency. This phenomenon is known as the constant phase angle (CPA) response in electrochemistry [1]. It can also be expressed in terms of the power law exponent appearing in the phenomenological Cole-Davidson expressions for the complex dielectric response, $\varepsilon^*(\omega)$. Apparently, many systems [2, 3] at ambient temperatures can be characterized by the exponent, $1>\beta>0$, in the Cole-Davidson expression, eq. (1) below, or the exponent, $\alpha = 1-\beta$, in the admittance representation.

This power law characteristic is often referred to as the "universal power law" frequency response [4, 5].

$$\varepsilon^*(\omega) \propto 1/(1+i\omega\tau)^{\beta} \qquad \text{eq. (1)}$$

where $\omega$, $\tau$, and, i, are the angular frequency, the time constant, and $\sqrt{-1}$, respectively.

Pure Debye response is characterized by, $\beta = 1$. It corresponds to a frequency response of a single, series R-C element [5]. At high frequencies, $\omega >> 1/\tau$, the capacitance part of this circuit is, effectively, shortened. The resistive component is then responding to the AC signal with a $0^0$ phase difference between the applied AC current and the resulting voltage drop. Using the complex permittivity representation, the high frequency response is expressed by separating the real part, proportional to $1/(1+(\omega\tau)^2)$, and imaginary part, proportional to $\omega\tau/(1+(\omega\tau)^2)$, of eq. (1), and noting that the real part of, $\varepsilon^*$ (pure capacitance), becomes negligible, while the imaginary (loss) part becomes inversely proportional to frequency, which corresponds to a constant conductivity in the admittance representation.

At very low frequencies, $\omega << 1/\tau$, the phase angle is $90^0$ as the capacitor has the dominant impedance to the flow of the applied AC current. This capacitive response is expressed through the finite and constant real part of, $\varepsilon^*$, while the imaginary, loss part, in eq. (1) becomes insignificant. It should be noted that, for the simple R-C Debye element, there is a sharp transition between the phase angle of $90^0$ at low frequencies to $0^0$ phase angle at high frequencies, at the $1/\tau$ characteristic frequency. In contrast to this characteristics of the phase angle change with frequency for the Cole-Davidson response with, $1>\beta>0$, shows the characteristic CPA plateau, as shown in fig. (2).

Many heterogeneous systems are characterized by the universal power law response with, $1-\beta \approx 0.6 - 0.7$ [2, 3]. Such systems can not be represented by simple series/parallel R-C equivalent circuits. We have suggested [6, 7] the use of a mixture of series and parallel R-C connections, which are inherent to arrays of resistors and capacitors placed at random. The power law characteristics, and the CPA, were shown to be the result of the logarithmic, series-parallel mixing of resistors' and capacitors' responses. The reasoning behind the series-parallel mixing was based on the "in-between" the arithmetic and the hyperbolic averaging expression, to model the mixed connectivity characteristics in random RC networks.

In a network having equal numbers of resistors and capacitors, the frequency was shown [7] to follow eq. (1) with, $\beta = 1/2$, corresponding to a $45^0$ CPA and power laws of $1/\omega^{0.5}$ for both the real ($\varepsilon'$) and the imaginary ($\varepsilon''$) parts of the complex dielectric constant ($\varepsilon^*$) at high frequencies.

At intermediate values of the ratio between the number of capacitors and resistors in the network, $1-\beta$ was proportional to this ratio. For example, a simulation of a network containing 60% capacitors and 40% resistors shows [7, Fig. 3] a $54^0$ phase plateau corresponding to $\beta = 0.4$ power in the Cole-Davidson expression .

It should be noted that the phase angle referred to in Fig. 2 below is the current-voltage phase angle, corresponding to the admittance representation. It is related to the conductivity power law exponent, or the logarithmic mixing power $\alpha$, applied to spatially disordered, series-parallel R-C nets, by $\alpha = 1-\beta$ [**6**–8]. The admittance phase angle is then proportional to $\alpha$ by, $\alpha(\pi/2)$, and $\alpha$ itself is given by the fraction of the dielectric phase in the binary mixture.
As noted above, the power law admittance response, in many experimental systems, stretches over a wide range of frequencies with, $\alpha = 1\text{-}\beta \approx 0.6 - 0.7$. While it can be argued that such values of, $\beta$ or $\alpha$, can characterize 3D systems at the percolation threshold for conductivity ($Pc \approx 1/3$), it is not clear why so many experimental systems should be "pinned" to the Pc composition in 3D. In particular, this percolation argument can not explain the *ac* response of 2D ionic conductors that, nevertheless, exhibit a universal, $\alpha \approx 0.6 - 0.7$, power law rather than the expected, $\alpha = 1/2$, exponent.
In the present work we are suggesting an explanation to the apparent universal value of $\alpha = 1-\beta$, the validity of which will not be limited to 3D systems at the percolation threshold composition.

<u>The model</u>

Let us first derive an expression for the equivalent resistance of a series-parallel resistive network, in which the individual resistors obey a uniform distribution of values. It had been argued before [9] that the equivalent value of a series-parallel network of elements should correspond to the geometric mean of the values of all its elements. The argument is based on the fact that the two limiting cases, the purely series and purely parallel connections, are related to arithmetic and hyperbolic means, which are characterized by the exponents, 1, and, -1, respectively. The series-parallel mixture corresponds to the "intermediate", geometric mean.
Given the argument above, the equivalent resistance of series-parallel random resistor network the individual resistances of which are uniformly distributed between a minimal resistivity, $R_0$, and a maximal resistivity, $R_1$, is the geometric mean, $R_g$, given by,

$$\ln(R_g) = \frac{1}{(R_1 - R_0)}\int_{R_0}^{R_1}\ln(x)dx = \frac{1}{R_1 - R_0}\left(R_1\ln(R_1) - R_0\ln(R_0)\right) - 1$$
$$= \ln\left(\frac{1}{e}R_1^{\frac{R_1}{R_1-R_0}}R_0^{\frac{-R_0}{R_1-R_0}}\right) \qquad \text{eq. (2)}$$

where, e, is the base of natural logarithms.
It should be noted that, following Hashin-Shtrikman model, the geometric mean derivation in the form of eq. (2) is usually applied to the conductivities of the elements, rather than to their resistivity values [9]. However, since the addition of resistors and conductors, in the two extreme cases of parallel and series connections, apply (in an opposite way) to the series-parallel mixing discussed above, this mixing can be used in the case of calculating the equivalent resistance of a collection of separate resistors, connected in a network.
It has also been shown that the series-parallel argument leading to the derivation of the geometric mean applies both to 2D and 3D systems [9], which is a significant result in the context of any model trying to explain why the 0.6-0.7 power law appears to apply in so many different systems. As been mentioned in the introduction, in percolation-based models the 3D case suggest the “pinning” to the Pc composition, while percolation in 2D can not explain the 0.6-0.7 power law, even in the case of “pinning”.
Eq. (2) is an extension of the well known result that the geometric mean of a continuous variable distributed uniformly within an interval {0, 1} is, $1/e \approx 0.37$.
Suppose now that a set of resistors spans over a wide range of values. Let the low resistivity limit be 100 ohm and the high resistivity limit be 100 Kohm.
Given that $R_1 >> R_0$, then, according to eq. (2), $R_1$, will dominate the expression, reducing the expression for the geometric mean to, $(1/e)R_1$. Note that an arithmetic mean would be, $(1/2)R_1$, given that the low (relative to $R_1$) value of $R_0$, can be ignored in the expression, $(R_1 – R_0)/2$. This is the result of the geometric mean “bias” towards small values.
The lower resistivity limit in this case, $R_0$ = 100 Ohm, is three orders of magnitude lower than, $R_1$. We therefore approach a limiting case of a uniform distribution of resistances between ~0 and 100 Kohm. The equivalent resistance of a random resistor network constructed from resistors of this distribution is then, $R_g$ = 37 Kohm, as argued above.
Using a statistical reasoning, lower values of resistances (higher conductivities) signify a higher probability of finding a unit conductor. By this reasoning, an equivalent resistance of, say, 50 Kohm in the uniform range of components’ values between ~ 0 and 100 Kohm (extreme values representing a binary-like split between unit value conductors vs. no conductors) would correspond to a probability of 0.5 for the bonds in the network to be occupied by unit conductors.
It is therefore suggested that it can be possible to map the position of the effective resistance value, $R_g$, in the continuous range of all resistivities between, $R_0$, and, $R_1$, into the probability of finding/not finding a unit conductor at a random location in the network. It characterizes an intermediate situation between the two limiting extremes, an entirely “nonconductive” network made of 100 Kohm resistors and, entirely conductive, all 100 Ohm elements network.
Suppose that $R_g$ turns out to be 100 Kohm. It is a reference state for 100% “holes” and 0% conductors. Let $P_h$ be the probability of finding a “hole” at a random position on the net. Then, $P_{cd} = 1 - P_h$, is the probability of finding a conductor at a random position on the net. For the 100 Kohm network, $P_h = 1$, and, $P_{cd} = 0$.
For a uniform distribution of resistances in the range between, $R_1$ and $R_0$ (where $R_1 >> R_0$) it is argued that the RC network considered here, can be represented by a partially populated network for which, $P_{cd} = 1 - P_h = 1 - 37/100 = 0.63$, fraction of bonds are conductive. The rest of bonds are not conductive.
In other words the geometric mean averaging has “biased” for low resistivity values (hence, for high conductivity values) relative to a possible arithmetic averaging that would have resulted in a 0.5 fraction of bonds being conductive.
To test the probabilistic model suggested above, R-C networks containing 50% resistors and 50% capacitors had been constructed. All capacitors are 1 nF. For resistors, a uniform distribution of resistors from 100 ohm to 100 Kohm has been used (see Fig. 1). This choice reflects the natural expectation for having a similar number of grains and gaps in granular or heterogeneous material, while the uniform distribution of resistivities could be rationalized as an expression of a random grain size distribution.
According to the argument above, a fraction, $0.5*P_{cd} = 0.315$, of all of the links in the network are conductive, as the “conductor – no conductor” transformation of the resistors' uniform distribution has reduced by the factor $P_{cd} = 0.63$ the maximal number of conductors (½ the total number of links in the network). On the other hand, according to the present model, capacitors always occupy 0.5 of network's links. Therefore, the fraction of capacitive bonds, relative to all "used" bonds (by capacitors and conductors) is, 0.5 / (0.5 + 0.315) = 0.61. This fraction should then be expressed as a $55^0$ ($0.61*\pi/2$) phase in the CPA region.
The results below (Fig. 2) clearly confirm the model suggested here.

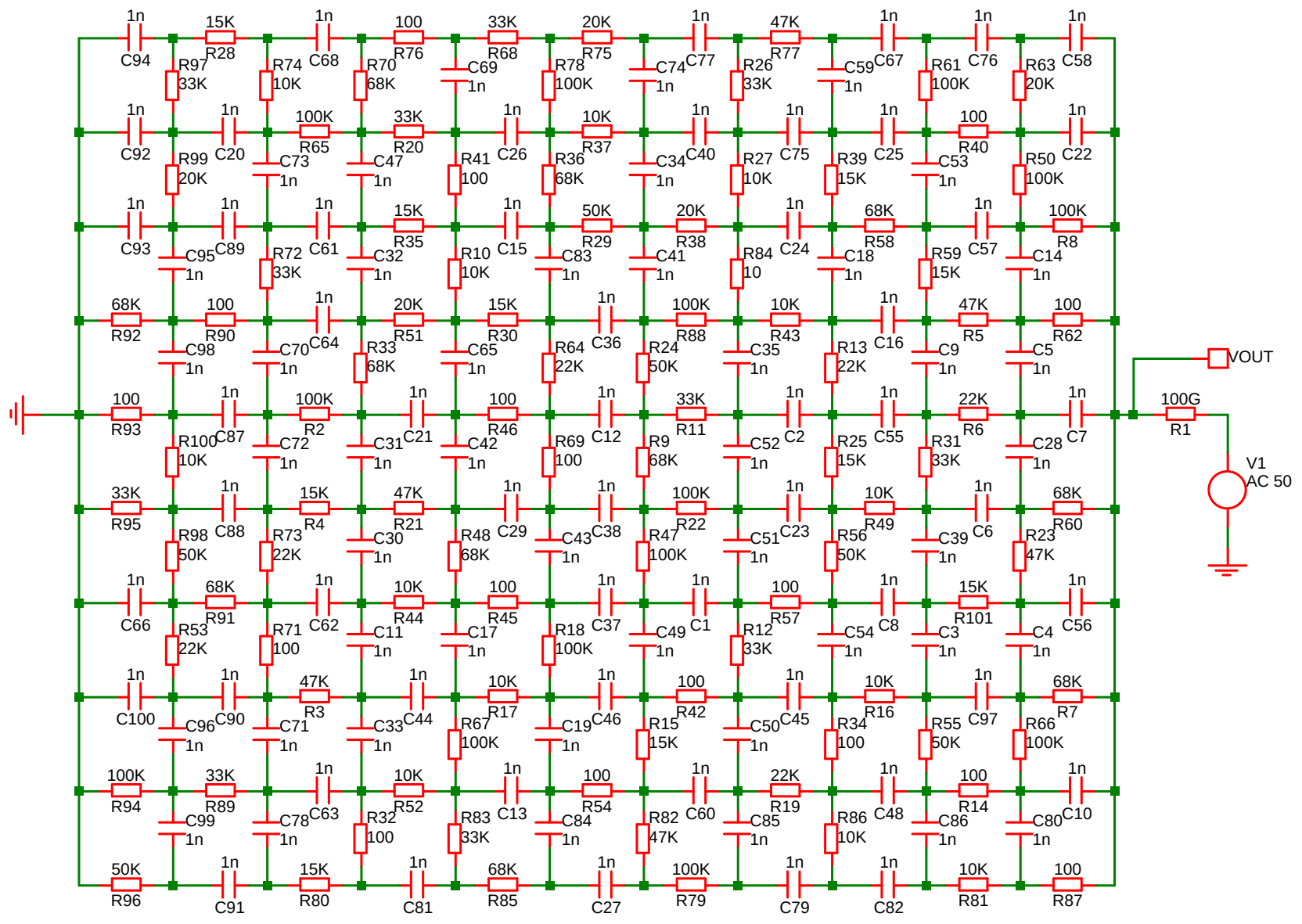

Fig. 1: Constant amplitude *ac* simulation of a random RC network having a uniform distribution of resistor values between 100 and 100000 ohms.

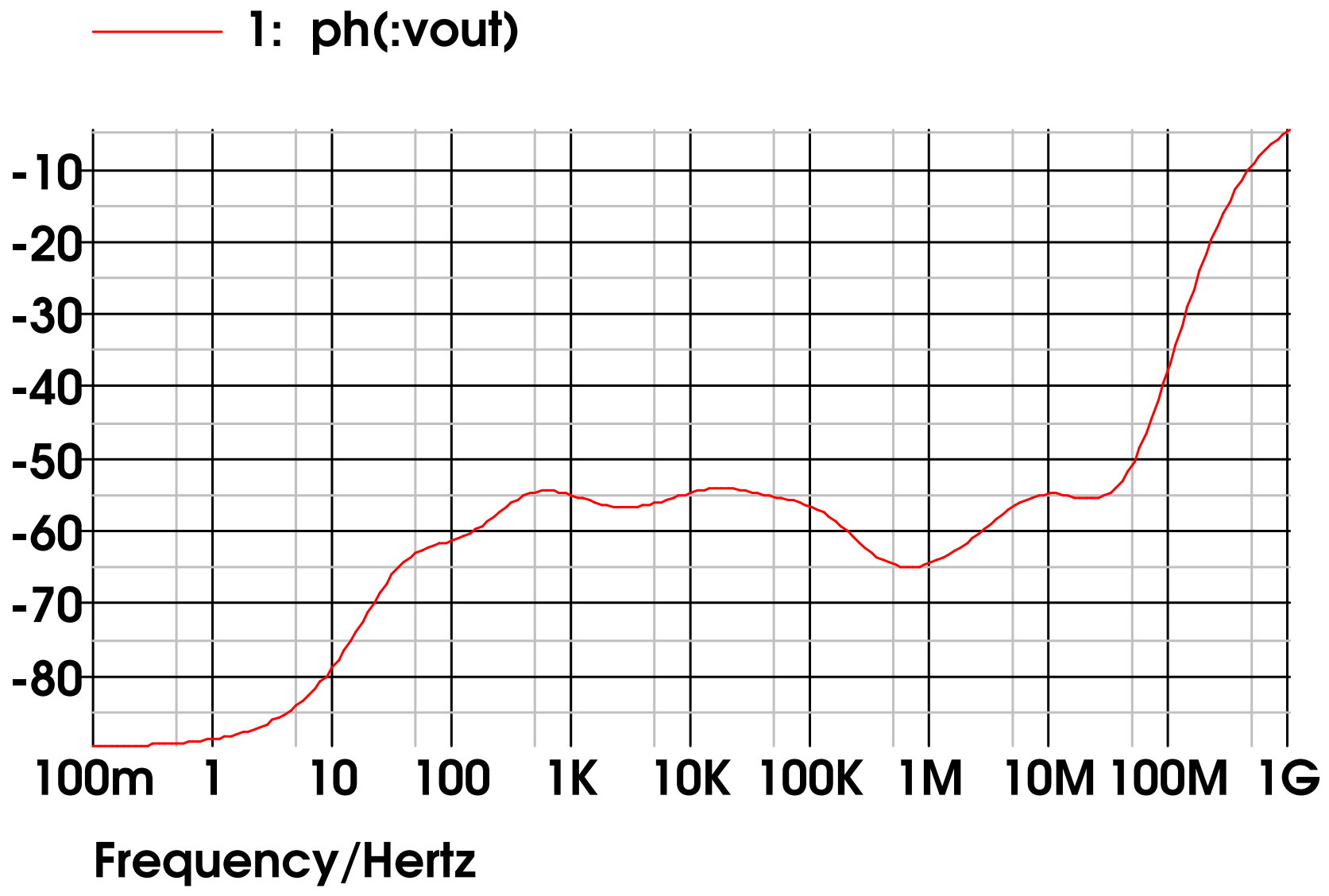

Fig. 2: The phase angle measured at the terminal "Vout" in Fig. 1 relative to the *ac* input to the random RC network.